%% file: sample-manuscript.tex
\documentclass{article}[10]

\usepackage{fullpage}
\usepackage{longtable}
\usepackage{lscape}
\usepackage{afterpage}
\usepackage{caption}
\usepackage{array}
\usepackage[utf8]{inputenc}
\usepackage{amsfonts}
\usepackage{amsthm}
\usepackage{subcaption}
\usepackage[english]{babel}
\usepackage{bm}
\usepackage[ruled,linesnumbered,vlined]{algorithm2e}
\usepackage{graphicx}
\usepackage{amssymb}
\usepackage{amsfonts}
\usepackage{blindtext}
\usepackage{verbatim}
\usepackage{float}
\usepackage{changepage}
\usepackage{listings}
\usepackage{xcolor}
\usepackage{tcolorbox}
\usepackage{parskip}
\usepackage{scrextend}
\usepackage{comment}
\usepackage{enumerate}
\usepackage{dsfont}
\usepackage{enumitem}
\usepackage{xr}
\usepackage{color}
\usepackage{hyperref}
\usepackage{pgfgantt}
\usepackage{pdflscape}
\usepackage{afterpage}
\usepackage{tikz}
\usepackage{pdflscape}
\usepackage{amsmath}
\usepackage[numbers]{natbib}
\usepackage{hhline}
\usepackage{booktabs}
\usepackage{mathtools}
\usepackage[nodisplayskipstretch]{setspace}
\usepackage{cleveref}
\usepackage[T1]{fontenc}

\usetikzlibrary{arrows}

\theoremstyle{plain}

\theoremstyle{definition}

\newcolumntype{L}[1]{>{\raggedright\let\newline\\\arraybackslash\hspace{0pt}}m{#1}}
\newcolumntype{C}[1]{>{\centering\let\newline\\\arraybackslash\hspace{0pt}}m{#1}}
\newcolumntype{R}[1]{>{\raggedleft\let\newline\\\arraybackslash\hspace{0pt}}m{#1}}

\begin{document}

\title{Job Recommender Systems: A Review}

\makeatletter
\renewcommand\@date{{%
  \vspace{-\baselineskip}%
  \large\centering
  \begin{tabular}{@{}c@{}}
    Corn\'{e} de Ruijt \textsuperscript{1} \\
    \normalsize c.a.m.de.ruijt@vu.nl
  \end{tabular}%
  \quad
    \begin{tabular}{@{}c@{}}
    Sandjai Bhulai \textsuperscript{1} \\
    \normalsize s.bhulai@vu.nl
  \end{tabular}

  \bigskip

  \textsuperscript{1}Faculty of Science\\ Vrije Universiteit Amsterdam\\ Amsterdam, the Netherlands \par

  \bigskip

  \today
}}
\makeatother

\maketitle

\begin{abstract}
This paper provides a review of the job recommender system (JRS) literature published in the past decade (2011-2021). Compared to previous literature reviews, we put more emphasis on contributions that incorporate the temporal and reciprocal nature of job recommendations. Previous studies on JRS suggest that taking such views into account in the design of the JRS can lead to improved model performance. Also, it may lead to a more uniform distribution of candidates over a set of similar jobs. We also consider the literature from the perspective of algorithm fairness. Here we find that this is rarely discussed in the literature, and if it is discussed, many authors wrongly assume that removing the discriminatory feature would be sufficient. With respect to the type of models used in JRS, authors frequently label their method as `hybrid'. Unfortunately, they thereby obscure what these methods entail. Using existing recommender taxonomies, we split this large class of hybrids into subcategories that are easier to analyse. We further find that data availability, and in particular the availability of click data, has a large impact on the choice of method and validation. Last, although the generalizability of JRS across different datasets is infrequently considered, results suggest that error scores may vary across these datasets. 
\end{abstract}

\section{Introduction}
\input{introduction.tex}

\section{Method and preliminaries}
\label{sec:methprelim}
\input{prelim.tex}

\section{Results}
\label{sec:results}
\input{method}

\section{Conclusion}
\label{sec:conclusion}
\input{conclusion.tex}
\bibliographystyle{plainnat}
\bibliography{references.bib}

\end{document}

%% file: introduction.tex
From the start of the commercialization of the internet in the late 1980s, the question was raised of how this technology could be leveraged in employee recruitment to enhance job seeker - vacancy matching. Even before the start of the world wide web, \citet{jose1990semantic} already proposed a system to match job seekers and jobs, which could \textit{``be consulted by Minitel, using telephone number 3615 and selecting the LM/EMPLOI service”}. I.e., the service allowed job seekers to send text messages in the form of search queries or their digital resume, over the telephone line, using a computer terminal called Minitel\footnote{https://en.wikipedia.org/wiki/Minitel}. The service would compare words in the query/resume to a knowledge base, which used a fixed job taxonomy to return a set of potentially interesting vacancies for the job seeker.

Although more than 30 years have passed since this early contribution, the usage of a fixed job taxonomy to extract information from a resume including ``branch of industry” (industry) and ``qualification” (skill) using \textit{``(a) dictionary specialized in the universe of employment”}, seems vaguely similar to LinkedIn’s query processing method \cite{li2016get}, which can be queried using your mobile phone\footnote{Parts between brackets are job attributes in the job taxonomy used by LinkedIn, to which queries are matched using dictionaries \cite{li2016get}}.  Of course, this is a very simplified view on reality: Vega’s 200 simultaneous Minitel connections could not have served the currently close to 750 million LinkedIn users worldwide \cite{aboutlinkedin}. Nonetheless, the problem of recommending the right job to job seekers remains as pressing as it was more than 30 years ago.

In this paper, we will provide an overview of the literature on job recommender systems (JRS) from the past decade (2011-2021). We will consider the different methods used in these systems, and consider these from a reciprocal, temporal, and ethical perspective. Furthermore, we will consider the influence of data availability on the choice of method and validation, and put extra emphasis on branching the large class of hybrid recommender systems used in this application domain. 

Our results suggest that JRS could benefit from a more application-oriented view: the reciprocal and temporal nature of JRS are infrequently discussed in the literature, while contributions that do consider these show considerable benefits. Furthermore, fairness is rarely considered in job recommender systems, and if it is considered, authors too often conclude that removing discriminatory features from the data is sufficient. However, recent scientific attention on fairness, in particular in candidate search engines, has introduced various metrics and algorithms, which we believe have the opportunity to be applied in the job recommender domain as well. In accordance with recommender system literature, deep language models have also been more frequently applied in the domain of job recommender systems. What remains somewhat unknown, is how well results on JRS generalize across datasets: the only study that considers this question shows that error metrics my vary greatly over different datasets.

This paper has the following structure. Section \ref{sec:methprelim} discusses some earlier literature surveys, thereby providing an argument for our interest in writing this review. Also, it discusses the method used for obtaining and selecting literature, and provides a brief discussion on some datasets and terminology we will frequently mention in this paper. Section \ref{sec:results} discusses our findings, in which Section \ref{subsec:classictax} classifies the job literature into the familiar recommender system taxonomy, while at the same time branching the large class of hybrid contributions. The remainder of Section \ref{sec:results} considers auxiliary topics such as the influence of data science competitions (Section \ref{sec:competitions}), validation of JRSs (Section \ref{sec:jrseval}), JRSs taking a temporal and/or reciprocal view (Section \ref{subsec:temprec}), ethical considerations in JRS (Section \ref{subsec:ethics}), and considers contributions discussing job search and recommendation at LinkedIn (Section \ref{subsec:linkedin}). Last, Section \ref{sec:conclusion} draws a conclusion and discusses directions for further research.

%% file: prelim.tex
\subsection{A brief discussion of previous surveys}
\label{subsec:prevsurvey}
Before we discuss the literature, it is useful to observe that in recent surveys on applications of recommender systems, job recommender systems and (more general) recommender systems in e-recruitment, are frequently not included. I.e., in the well-cited review on applications of recommender systems, \citet{lu2015recommender} do not mention the application area of e-recruitment, the same holds for the earlier review by \citet{felfernig2013toward}. Also, although most papers on neural networks in job recommender systems were published after 2018, the survey on (deep) neural networks in recommender systems (including a section on application areas) also neglects this application \cite{batmaz2019review}. From the HR perspective, job search and recommendation are also not always mentioned as an application area, as opposed to candidate selection, while in the end these systems do determine who will be in the applicant pool in the first place \cite{strohmeier2013domain}. 

One possible explanation could be that, from a technical perspective, the problem of job search and job recommendation is little different from a general information retrieval/recommendation task. Job seekers frequently use `general-purpose’ search engines and online social networks to search for jobs (e.g., \cite{jansen2005using, de2016predicting,koch2018impact}). Furthermore, many job recommender systems we will discuss in this paper could very well be used in other application areas (and vice versa). Nonetheless, we will argue that factors such as the large amount of textual data, the reciprocal and temporal nature of vacancies, and the fact that these systems deal with personal data does require a tailored approach, and the sheer volume of contributions make it clear that this application area should not be neglected. 

Previous surveys on job recommender systems, which consider JRS contributions before 2012, include \citet{al2012survey} and \citet{siting2012job}, though especially the latter survey is very limited in scope. More recent is the survey on recommender systems in e-recruitment by \citet{freire2020recruitment}. Although our work has some overlap, we especially wish to address some of the limitations of the work by Freire and de Castro in this paper. 

Even though the work by Freire succeeds in collecting a substantial number of contributions in the JRS application domain, they seem to fail to properly classify these contributions, making it difficult to see patterns in this literature. A clear example of this is that approximately 20\% of the contributions discussed in their paper is labeled as hybrid, whereas another 33\% is being labeled as ``other”. Although the reader would later find that the ``other" category includes for 25\% contributions using (deep) neural networks, this still leaves a large number of contributions with an unsatisfying label. Furthermore, as shown by \citet{batmaz2019review}, there is a considerable development within the class of (deep) neural networks applied to recommender systems, which we also find in job recommender systems. This aspect is neglected by Freire and de Castro.  

The classification given by Freire and de Castro is understandable, given that so many contributions use mixtures of collaborative filtering and content-based techniques, and given that these are presented by the contributions themselves as hybrids. However, these labels do not provide much insight into what these contributions actually entail. Furthermore, \citet{freire2020recruitment} focus solely on methods and validation, whereas we, among other subjects, will also take into consideration ethical considerations. We will also put special emphasis on job recommender systems which, often successfully, take into account the reciprocal and temporal nature of job recommendations.

\subsection{Preliminaries}
As we will discuss in Section \ref{sec:competitions}, many contributions use one of the data sources made available through data science competitions for training and validating job recommender systems. Most noticeably, these include the \textit{RecSys 2016} and \textit{RecSys 2017} competitions (\cite{abel2016recsys}, and \cite{abel2017recsys} respectively), using a dataset from the job board Xing\cite{xing2021}, and the \textit{CareerBuilder 2012} Job Recommendation Challenge \cite{careerbuilderkaggle}, which was hosted by CareerBuilder on Kaggle\cite{careerbuilder2021,kaggle}. All three datasets contain data with respect to candidate profiles, vacancies and online interaction between the two. Another resource often used is the Occupation Information Network (\textit{O*NET})\cite{onet2021}, an English-based job ontology that is frequently used in knowledge-based job recommender systems (see Section \ref{subsubsec:knowledgebasedjrs}). 

We will use the terms \textit{vacancy}, \textit{job posting}, and \textit{job} somewhat interchangeably throughout this paper to represent the \textit{item} in the classical recommender system setting, whereas job seekers are considered as \textit{users}. Although, as in the early paper by Vega\cite{jose1990semantic}, job seekers are still often described by their resumes, some current e-recruitment systems allow for descriptions that move beyond self-descriptions of one's professional self. Here one should think of the social connections one can observe on (professional) social networks. When we speak of \textit{resumes}, \textit{CVs}, \textit{user profiles}, or \textit{job seeker profiles}, we assume these are synonyms and may contain additional information (such as social relations) beyond the self-description. Last, we will sometimes speak of ``textbook" or ``off-the-shelf", by which we mean methods one can find in popular machine learning/pattern recognition textbooks such as by \citet{bishop2006pattern} and \citet{flach2012machine}, or \citet{aggarwal2016recommender} in the case of commonly used recommender systems. 

\subsection{Method}
To obtain the set of contributions we will discuss in this paper, the following method was employed. English contributions published between January 1st 2011 and January 1st 2021, published as an article in a scientific journal or in conference proceedings were searched using Google Scholar, using key phrases \textit{job recommender systems}, \textit{job recommendation}, and \textit{job matching}, where additionally we replaced ``job" by ``occupation" in these phrases. Furthermore, a forward and backward search was applied, which after reviewing the titles and abstracts resulted in 192 contributions. 

Next, contributions on career trajectory recommendation were omitted, that is, recommender systems that only recommend a job type (such as ``data scientist"), but not a specific occupation at a given organization at a specific time. Furthermore, also papers on employee selection, recommender systems specifically for freelancers, and candidate recommender systems (i.e., recommending candidates for open vacancies) were filtered out. We further removed contributions without (some form of) empirical validation of their recommender system. The resulting set contained 87 contributions, which we will discuss in this paper.

%% file: method.tex
\subsection{Methods in job recommender systems}
\label{subsec:classictax}
As discussed in Section \ref{subsec:prevsurvey}, although the work by \citet{freire2020recruitment} is considerable, our critique lies in how they classify the different job recommender systems.  As is common in the literature (e.g., \cite{aggarwal2016recommender}), these are split into Content-Based Recommender Systems (CBR), Collaborative Filtering (CF), Knowledge-based Recommender Systems (KB), and Hybrid Recommender Systems (HRS). However, given that so many contributions are hybrid, we will further make a distinction between \textit{monolithic} and \textit{ensemble} hybrid recommender systems \cite[p. 200]{aggarwal2016recommender}\footnote{\cite{aggarwal2016recommender} also considers the \text{mixed design}, which we did not find in the JRS literature, and is therefore excluded}, and split the class of ensemble hybrids using the classification introduced by \cite{burke2002hybrid}. The interpretation of monolithic and ensemble hybrids, and the different ensemble hybrids, will be discussed in Section \ref{subsec:hybridjrs}.

\begin{table}
\tiny
\centering
\caption{Contributions per year per JRS category}
\label{tbl:litoverview}
\resizebox{\textwidth}{!}{%
    \begin{tabular}{ll|lllllllll}
         &          & \textbf{Method} &                                 &                 &             &           &           &                    &      &           \\
         &          &     &                   & \textit{Ensemble hybrid} &             &           &           & \textit{Monolithic hybrid}  &      &         \\
    Year & Dataset & CBR    & CF            & Cascade         & Feature aug.& Switching & Weighted  & MM-SE              & DNN  & KB  \\
    \hline
    2011 &   O     &        &                                 &                 &             &           &           &  \cite{paparrizos2011machine}                  &      &      \\
    2012 &   O     &   \cite{kessler2012hybrid}     &                                 &                 &             &           &           &                    &      &      \\
    2013 &    O      &        &                                 &   \cite{lu2013recommender}              &  \cite{diaby2013toward}           &           & \cite{bostandjiev2013linkedvis, hong2013job}          &  \cite{wang2013time}                  &      &      \\
    2014 &    O      &        &                                 &                 &  \cite{diaby2014exploration, diaby2014taxonomy,faliagka2014line,gupta2014applying,malherbe2014field}           &           &  \cite{heap2014combining}         &      \cite{poch2014ranking}              &      &      \\
         &  CB12        &        &                                 &                 &             &           &           &                    &      & \cite{guo2014analysis}     \\
    2015 &   O       &   \cite{chenni2015content}     &                                 &                 &             &           &           &                    &      & \cite{kmail2015matchingsem,kmail2015automatic}    \\
    2016 &   O       &  \cite{domeniconi2016job}      &                                 &   \cite{gui2016downside}              &    \cite{lee2016job}         &  \cite{schmitt2016matching}         &   \cite{lin2016machine,zhang2016glmix}        &                    &      &  \cite{li2016get}    \\
         &   RS16       &        &   \cite{ahmed2016user,liu2016hybrid}                              &  \cite{de2016scalable,honrado2016jobandtalent,mishra2016bottom,pacuk2016recsys,wang2016analysis,xiao2016job}               &             &   \cite{zhang2016ensemble}        & \cite{polato2016preliminary,leksin2016job,zibriczky2016combination,carpi2016multi}          &                    &   \cite{liu2016temporal}   &      \\
    2017 &  O        & \cite{bansal2017topic}       & \cite{reusens2017note,lee2017exploiting}                                &  \cite{borisyuk2017lijar}               &             &           &  \cite{tran2017comparison}         &  \cite{chen2017hybrid,dong2017job,schmitt2017language,yang2017combining}                  &      & \cite{zaroor2017hybrid,zaroor2017jrc}     \\
         & CB12         & \cite{shalaby2017help}       &                                 &                 &             &           &           &                    &      &      \\
         &  RS17        &        &                                & \cite{lian2017practical,volkovs2017content,yagci2017ranker,sato2017exploring}                &   \cite{guo2017integration}          &           &   \cite{leksin2017combination,bianchi2017content}        &                    &      &  \cite{ludewig2017light}    \\
    2018 &   O       & \cite{janusz2018match,valverde2018job}       &                                 &                 &             &           &           &  \cite{maheshwary2018matching,dave2018combined,jiang2019user,xu2018matching}                  & \cite{qin2018enhancing,zhu2018person}     &  \cite{reusens2018evaluating}    \\
         &   CB12       &        &                                 &                 &             &           &           &                    &      & \cite{almalis2018constraint}     \\
         &  RS16        &        &                                 &                 &             &           &           &  \cite{chen2018tree}                  &      &      \\
    2019 &   O       & \cite{lacic2019should}       &                                 &                 &             &           &           &                    &  \cite{luo2019resumegan,bian2019domain,le2019towards,nigam2019job,yan2019interview}    & \cite{gutierrez2019explaining,martinez2019novel,shishehchi2019jrdp,rivas2019hybrid}     \\
    2020 &   O       & \cite{gugnani2020implicit,chou2020based}       &                                 &                 &             &           &           &                    & \cite{bian2020learning,jiang2020learning,qin2020enhanced}     &      \\
         &   CB12       & \cite{mpela2020mobile}       &                                 &                 &             &           &           &                    &      &      \\
         &   CB12, RS17, O       &        &  \cite{lacic2020using}                              &                 &             &           &           &                    &      &      \\
    \bottomrule
    \end{tabular}}
    \\
    
    \medskip\tiny
    Dataset abbreviations: Career Builder 2012 (CB12), RecSys 2016 (RS16), RecSys 2017 (RS17). `O' implies another dataset was used in these papers.
    
\end{table}

\subsubsection{Content-based JRS}
Content-based recommender systems (CBRs) in the context of JRS are models which, to construct a recommendation, only use a semantic similarity measure between the user profile and the set of available vacancies. I.e., the semantic similarity is used as a proxy for estimating the relevance of each vacancy to the job seeker. In CBRs, one creates vector representations of the vacancy and user profile in an unsupervised way, i.e., the dimensions of these representations may not have an intuitive interpretation. Many authors use Bag of Words (BoW) with TF-IDF weighting \cite{mpela2020mobile, kessler2012hybrid, chou2020based, domeniconi2016job,chenni2015content}, though also Latent Dirichlet Allocation is used \cite{bansal2017topic}, and the more recent word2vec \cite{gugnani2020implicit, valverde2018job, janusz2018match}. Interestingly, CBR contributions have been relatively stable in the past 10 years, but also, they were not part of the top contributions during the 2016 and 2017 RecSys competitions (see Table \ref{tbl:litoverview}).

Perhaps the main challenge in content-based JRS is that, in the words of \citet{schmitt2016matching}, \textit{``Job seekers and recruiters do not [always] speak the same language”}. I.e., job seekers and recruiters may use a different terminology to describe jobs, knowledge, or skills. As a result, two entities having the same meaning may end up in different vector representations when described by the recruiter or by the job seeker. Unfortunately, this discrepancy is often neglected in content-based JRS.

\subsubsection{Collaborate filtering JRS}
In collaborative filtering (CF), recommendations are based solely on behavioral data, typically stored in a user $\times$ items rating matrix. In the e-recruitment setting, like in the e-commerce setting, this matrix is usually filled with click behavior (e.g., the rating matrix equals 1 if the job seeker (row) clicked on a vacancy (column) to see the vacancy details, 0 otherwise). Though, in case such interaction data is missing, also the sequence of previous job occupations can be used to fill the rating matrix (e.g., \cite{lee2017exploiting}). The latter case does require jobs to be categorized, such that they can serve as discrete items in the rating matrix. For simplicity, we will refer to the entries in the rating matrix as ratings, irrespective of the exact type of behavioral feedback. 

The literature commonly distinguishes between two types of CFs: memory-based CF and model-based CF \cite{aggarwal2016recommender}. 
In memory-based CF, recommendations are created using a $K$-nearest neighbor (KNN) type of approach. That is, for some user $u$, one tries to find either $K$ users similar to $u$ (user-based CF), or $K$ items similar to the items $u$ has already ranked (item-based CF). Here, the similarity is always based on the entries of the rating matrix. Contributions using textbook CF methods include \citet{lee2017exploiting, reusens2017note, ahmed2016user, liu2016temporal}, where the latter two applied it to the RecSys 2016 dataset. \citet{lacic2020using} compare several auto-encoders to encode user interactions, based on which similar users are determined. 

Model-based CF attempts to fill the missing values in the rating matrix using regression-based models, based solely on the rating matrix. However, likely due to the large amount of textual data, which can ideally function as features in such regression models, we are not aware of any studies using model-based CF in job recommender systems. On the other hand, although we classified regression models using textual data as hybrid, it should be noted that these models have a strong CF flavor. Even though textual features are used, their weights in the regression model are determined by behavioral feedback. Hence, even though not only behavioral data is used, still the resulting recommendation is largely determined by what vacancies similar job seekers have interacted with. Only the definition of `similar’  has changed, as it now partly includes features based on content.

\subsubsection{Hybrid JRS}
\label{subsec:hybridjrs}
As discussed earlier, hybrid recommender systems combine several models into one recommender system. Our aim here is to split this group into smaller, more homogeneous methods. To do so, we follow \citet[pp. 199-204]{aggarwal2016recommender}, who split the hybrid recommender systems into those having a \textit{monolithic}, or \textit{ensemble} design. A monolithic design refers to hybrid recommender systems in which it is not possible to extract one component of the hybrid, and build recommendations based on this component independent of the other components in the hybrid, without altering the algorithm. On the contrary, ensemble methods do allow splitting the hybrid recommender system into at least two components that can (independently) be used for recommendation. I.e., the ensemble may consist of some models which by itself may be monolithic.

\paragraph{Model-based methods on shallow embeddings}
In the class of monolithic designs, we again consider two classes. We will refer to the first class as \textit{model-based methods on shallow embeddings} (MM-SE). These approaches commonly use off-the-shelf machine learning methods such as support vector machines \cite{poch2014ranking}, naive Bayes \cite{paparrizos2011machine}, or gradient boosting \cite{yang2017combining, xu2018matching}, though also convolutional neural networks (CNNs) \cite{maheshwary2018matching} are used. What unifies these contributions, is that the textual data is mapped to a vector space using linear or small order transformations. Although we will leave the definition of shallow or small order somewhat vague on purpose, these embeddings do have in common that this order is explicit in the method, whereas in deep embeddings this could be arbitrarily large. These shallow embeddings include TF-IDF weighted document representations \cite{xu2018matching}, representations based on a predefined job classification \cite{yang2017combining, paparrizos2011machine}, word2vec \cite{maheshwary2018matching}, and a probabilistic stacked denoising auto-encoder \cite{chen2017hybrid}. 

Also graphical models are included in the MM-SE class of monolithic hybrids. \citet{dave2018combined}, map jobs and skills to two matrices sharing the same latent space. Although the mapping is trained independently from O*NET, the resulting vector representations showed considerable resemblance to the O*NET ontology. \citet{jiang2019user}, extract a large number of features from their dataset of the vacancies, job seeker profiles, and their interactions, to train a graphical model in which clicks are predicted based on estimates of whether a (job seeker, vacancy) pair is of some latent type $z$. They also make use of O*NET and use bin-counting (i.e., \cite[Ch. 5]{zheng2018feature}) to obtain a numeric representation of some categorical features. Although the model provides an explicit click probability, as opposed to only predicting the appropriate ranking of items, the paper does mainly compare its approach with learning to rank models such as AdaRank and RankBoost, where the model is found to be favorable.

Three contributions model the job recommendation problem with a somewhat different objective function, though, we still label these as MM-SE. \citet{dong2017job} and subsequent work \cite{chen2018tree} propose an MM-SE monolithic hybrid. Contrary to the approaches discussed so far, they consider the problem as a reinforcement learning problem. Given a successful or failed recommendation, a reward function is updated, taking as input the representation of a job seeker and a vacancy. To speed up computations, vacancies are represented in a binary tree.  \citet{wang2013time} not only consider what job should be recommended, but also when this recommendation should be made, based on estimations of when a job seeker will switch jobs.

\paragraph{Deep neural networks}
Apart from shallow embeddings, also deep representations have become common strategies, accounting for approximately 50\% of all contributions in 2019 and 2020 (see Table \ref{tbl:litoverview}). Since this class consists of only deep neural networks (DNNs), we will also refer to this class by DNN. Many studies follow a similar approach as language models such as BERT \cite{devlin2018bert} or ELMo \cite{Peters:2018}, and extend on these embeddings by adding additional hierarchical attention networks \cite{nigam2019job, qin2018enhancing}. Also embeddings based on CNNs are common \cite{le2019towards}, or mixtures of the above approaches \cite{luo2019resumegan, bian2019domain, bian2020learning, jiang2020learning, yan2019interview}. Siamese neural networks have also been used for this purpose \cite{schmitt2017language}. 

The increasing usage of DNNs in recommender systems is not limited to recommending jobs, but holds for recommender systems in general \cite{batmaz2019review}. Although it is too soon to draw firm conclusions, the absence of MM-SE contributions in the last two years is somewhat striking, giving that these were quite common in 2017 and 2018. Furthermore, some contributions did benchmark their proposed DNN with models that we may label MM-SE, e.g., using word2vec or BoW to obtain vector representations, and found the DNN to be favorable \cite{qin2018enhancing, nigam2019job, zhu2018person}. Also, in some  DNN contributions, the neural network on top of deep job/user/session embeddings is compared to other commonly used machine learning algorithms, such as gradient boosted decision trees (GBDT), with the latter using the same job/user/session embedding as the proposed DNN. In such comparisons, DNNs are also found to outperform other machine learning models on the same embeddings \cite{luo2019resumegan, jiang2020learning}.

To what extend this argument is completely satisfying remains somewhat unknown. If we view the architecture on top of initial job/user/session embeddings as automated feature engineering, then comparing these models with models without additional feature engineering is perhaps not completely fair. Moving away from the application of job recommender systems, DNNs also have not always been outperforming gradient boosting or ensemble techniques in recent data science competitions, in which participants put more effort on feature engineering for these boosting/ensemble models \cite{jannach2020deep}.

\paragraph{Ensemble hybrids}
Following \citet{burke2002hybrid}, we split ensemble hybrids into four classes. \textit{Cascade hybrids} refine the recommendation given by another recommender system. These commonly include (gradient) boosting models \cite{mishra2016bottom, lian2017practical, pacuk2016recsys, wang2016analysis, xiao2016job, volkovs2017content,sato2017exploring}, where in particular XGBoost\cite{chen2016xgboost} is popular. It should be noted that all these contributions were competitors in either the 2016 or 2017 RecSys competitions,
with \cite{xiao2016job} and \cite{volkovs2017content} being the winning contributions to the 2016 and 2017 competitions respectively. The problem has also been addressed from a learning to rank perspective, using LambdaMART \cite{honrado2016jobandtalent}. Besides boosting, also refinements using common CBR or CF methods have been proposed \cite{de2016scalable, yagci2017ranker, lu2013recommender}

\citet{gui2016downside} propose the so-called benefit-cost model, a re-ranking procedure for recommender systems that takes into account downside in recommender systems, and which is validated on a job recommender problem. To our knowledge, this contribution is the only contribution that raises the question of whether in some scenarios a recommendation should be given at all. Ill fitted recommendations could hurt the trust job seekers have in the recommender system, making them decide not to use it \cite{laumer2018job}. 

\citet{borisyuk2017lijar} refine an earlier job recommendation by predicting the number of applications to each vacancy, and pushing less popular vacancies to the top of the recommendation list. The idea behind this strategy is that, contrary to the e-commerce recommender setting, it can be hurtful to have too many applicants on a single vacancy. Recruiters will have to spend at least some time on each application to evaluate the fit and communicate the result of this evaluation to the  job seeker (e.g., communicate a rejection or invite the job seeker for an interview). Furthermore, \citet{borisyuk2017lijar} also find that pushing less popular vacancies does increase the number of applications to those vacancies, hence causing a better spread of the applicants over the vacancy portfolio, though this is conditioned on whether the less popular vacancies are sufficiently relevant.

Similar to cascade hybrids is the \textit{feature augmentation} class, in which case the result of the previous recommender system in the sequence is not so much refined, but simply used as an input for the next model. The somewhat similar approaches by \citet{diaby2013toward, diaby2014exploration, diaby2014taxonomy}, and \citet{malherbe2014field, faliagka2014line, gupta2014applying, guo2017integration} use such an approach, in all cases applying off-the-shelf machine learning methods on top of an initial CBR. 

In \textit{weighted hybrids}, the output of the separate models is combined using some (possibly non-linear) combination of the predicted scores. Commonly, CBR and model-based CF are in this way combined \cite{polato2016preliminary,leksin2016job,leksin2017combination,tran2017comparison,heap2014combining,carpi2016multi,bianchi2017content}. Others combine a large number of textual and behavioral features \cite{zibriczky2016combination,zhang2016glmix,bostandjiev2013linkedvis}, which we consider an ensemble hybrid as some of these features may be used for job recommendation by themselves. The output of several off-the-shelf machine learning methods has also been used for this purpose \cite{lin2016machine}. 

A common problem in CF is the cold-start problem: new users/items have not given/received any behavioral feedback yet, making CF-based recommendations difficult. Although multiple hybrid approaches can be used to resolve this problem, perhaps the most direct approach is to use \textit{switching hybrids}. In JRS, this most often implies that the recommender system uses CF by default. However, if an item or a user has insufficient data, the recommender system switches to CBR. Contributions using such approach with off-the-shelf CF and CBR methods include \cite{zhang2016ensemble}, and \cite{schmitt2016matching}.

\subsubsection{Knowledge-based JRS}
\label{subsubsec:knowledgebasedjrs}
Although \citet{aggarwal2016recommender} define knowledge-based recommender systems as recommender systems having the conceptual goal to  \textit{``Give me recommendations based on my explicit specifications of the kind of content (attributes) I want”}, we will rather use the definition given by \citet{freire2020recruitment}, who define it as \textit{``[Recommender systems]  which rely on deep knowledge about the product domain (job) to figure out the best items (job vacancies) to recommend to a user (candidate)"}. In job recommender systems, this often implies that both job and candidate profiles are mapped to some predefined job ontology (i.e., the knowledge base), after which the two are matched. 

A common strategy to generate job recommendations is then to compute the similarity between the candidate profile and vacancy in the ontology space \cite{rivas2019hybrid,kmail2015matchingsem,martinez2019novel,kmail2015automatic,zaroor2017hybrid,zaroor2017jrc,guo2014analysis,shishehchi2019jrdp,almalis2018constraint,ludewig2017light}, where the overlap can be computed by for example the Jaccard index (\cite{guo2014analysis}). Although one can imagine that the construction of such ontologies can take considerable effort, they have been used successfully in practice by, for example, LinkedIn \cite{li2016get} or Textkernel \cite{textkernel2021}. Nudging users to complete their profile by recommending skills from the ontology has also been shown to be a successful strategy to improve such recommendations \cite{bastian2014linkedin}. Mapping jobs and candidate profiles to a shared ontology also provides a solution to the discrepancy between how job seekers and recruiters define jobs without requiring behavioral feedback.

An advantage of using job ontologies is that this also simplifies the implementation of keyword-based search engines and simplifies filtering. \citet{gutierrez2019explaining} consider such an approach, in which an interactive recommender system was built on top of the knowledge-based recommender ELISE \cite{elisejob2021}, where users are able to filter jobs based on travel distance or the type of contract. The recommender system also explains why the recommendation is (not) given, e.g., by indicating that a required skill is not mastered by the job seeker. Participants in the study indicated that the tool allowed (citing the authors) \textit{``greater autonomy, a better understanding of needed competencies and potential location-based job mobility”}. 

\citet{reusens2018evaluating} compare several CF approaches with keyword search in terms of recall and reciprocity, both for job recommendation and job seeker recommendation. With respect to job recommendations, although traditional CF approaches performed acceptably in terms of recall, they were inferior to keyword search in terms of reciprocity. The authors did find that using a `reversed' rating matrix for job recommendation (that is, building a rating matrix on whether recruiters saved job seeker profiles for specific vacancies) improved reciprocity even beyond keyword search, though with a trade-off in terms of recall. 

\subsection{Competitions}
\label{sec:competitions}
Competitions have played an important role in the job recommendation literature. These competitions are settings in which organizations share a dataset, an objective, and an error measure with respect to this objective, typically via a competition platform such as Kaggle\cite{kaggle}. Teams can enroll in these competitions, and in the case of job recommender competitions, the goal for each team is to construct a job recommendation for a set of hold-out users. As already hinted by Table \ref{tbl:litoverview}, to our knowledge three competitions have had a considerable impact on the job recommender literature: 1) The 2012 Careerbuilder Kaggle competition \cite{careerbuilderkaggle}, 2) the RecSys 2016 competition \cite{abel2016recsys}, and 3) the RecSys 2017 competition \cite{abel2017recsys}. The latter two both used a dataset from the German job board Xing \cite{xing2021}. 

Besides being used for contributions to the competitions themselves, the datasets are also commonly used after completion of the competition, e.g., to train and validate job recommender systems when no dataset is available. Given that approximately 32\% of all job recommender system contributions use a dataset originating from one of these contributions, these datasets have had a considerable influence on the job recommender literature. Although many models proposed for the RecSys 2016 and 2017 competitions were cascade or weighted hybrids (and in particular gradient boosted trees were quite successful \cite{xiao2016job,volkovs2017content}), there is no free lunch: all contributions show that a considerable time and effort was spent on constructing useful features to represent job seekers, jobs, and their interaction. 

We will emphasize two more observations with respect to competitions. First, the 2012 CareerBuilder dataset contains user interactions, hence could be used to evaluate a collaborative or hybrid recommender system. However, it is frequently used to evaluate content-based recommender systems (see Table \ref{tbl:litoverview}). Second, to our knowledge, only \citet{lacic2020using} compare their proposed recommender system over multiple datasets: the Careerbuilder dataset, the 2017 RecSys dataset, and a private dataset originating from the student job portal Studo Jobs\cite{studo2021}. What is striking, is that the performance of the different models differs considerably across datasets, with no unanimous agreement across data sets and across error measures which model should be preferred. I.e., the results are rather dataset dependent. However, the authors do find that from the set of models they consider, using variational auto-encoders to embed user sessions gave overall the best performance.

\subsection{Validation}
\label{sec:jrseval}
\paragraph{Validation when lacking job seeker - vacancy interaction}
Many authors state in their introduction that there is a vast amount of online data available, commonly followed by a reference to the current LinkedIn user count \cite{aboutlinkedin}. Although this may be true in terms of vacancies and job seeker profiles, researchers do not always have access to interactions between the two. Naturally, this also impacts the type of methods and validation that is used.

In case interaction data is lacking, authors propose several strategies to still be able to evaluate their recommendations. One of these is to use one of the competition datasets for validation and training. As already discussed in Section \ref{sec:competitions}, approximately 32\% of all contributions use one of the competition datasets, but from Table \ref{tbl:litoverview}, we also find that these were used after the competitions had finished. Expert validation is also used frequently for validation. During such expert validation, the quality of recommendations is inquired by a group of `experts’, which may be the researchers themselves, HR/recruitment experts, or sometimes students (e.g., \cite{maheshwary2018matching}). Although the choice for expert validation is rarely discussed, we do find that for CBR and KB job recommender systems, approximately half of the contributions use expert validation  \cite{kessler2012hybrid,kmail2015matchingsem,chenni2015content,kmail2015automatic,bansal2017topic,zaroor2017hybrid,zaroor2017jrc,janusz2018match,valverde2018job,martinez2019novel,shishehchi2019jrdp,gugnani2020implicit}.   

Another way to obtain behavioral feedback when interaction data is lacking, is by using previous $N$ jobs in the job detail section of a resume to predict the $N+1$-th job. This does come with the challenge of rightfully defining a job. As mentioned by \cite{deruijt2021comparison}, job seekers may have different interpretations of a ``job". That is, if the definition is too specific, each job would have too few observations for inference. Whereas if the definition is too broad, the recommendation may be accurate, but not precise, and therefore not useful. Furthermore, many details about the job may be missing from the job history, and job seekers tend to indicate their jobs at different levels of granularity \cite{deruijt2021comparison} (e.g., are multiple related positions at the same firm one or multiple ``jobs"?). 

A somewhat interesting observation with respect to the best performing contributions during the RecSys 2016 and 2017 competitions is the fact that none of the best performing contributions were CBRs. This may be because interaction data was given, hence, why not use it? However, also in other hybrid recommender systems, incorporating behavioral feedback is found to improve the recommender system, even if one has access to a well-defined and validated job ontology \cite{li2016get}.

\paragraph{Choices in negative sampling}
In case interaction data is used for validation, one mostly only observes those jobs that the job seeker has positively interacted with. I.e., the job recommender system problem is most often modeled as an unary classification problem, and therefore training requires defining a set of negative samples. We consider two types of behavioral feedback, which commonly lead to somewhat different approaches in negative sampling. We consider using previously held jobs in the resume as positives as Type 1 behavioral feedback, whereas Type 2 behavioral feedback is obtained from observed online job seeker - vacancy interaction. 

For Type 2 behavioral feedback, common strategies for defining negatives include using shown but skipped items \cite{honrado2016jobandtalent,jiang2019user,le2019towards,liu2016temporal,mishra2016bottom,nigam2019job,pacuk2016recsys,polato2016preliminary,qin2020enhanced,xiao2016job,yang2017combining,zhang2016glmix,zibriczky2016combination,guo2017integration}, picking negative samples at random (not per se uniform) \cite{bian2019domain,yagci2017ranker,yan2019interview,leksin2016job,dave2018combined}, replacing the job (but not the candidate and further context) at random \cite{zhu2018person}, using vacancies of which the vacancy details were shown, but did not lead to an application \cite{qin2020enhanced,qin2018enhancing}, or if the method allows for sparse matrices (such as in some matrix factorization methods): using all possible vacancy-user interactions \cite{leksin2017combination,poch2014ranking,lee2016job,lacic2020using,bostandjiev2013linkedvis,reusens2017note,ahmed2016user,lee2017exploiting,liu2016hybrid,carpi2016multi,bianchi2017content,ludewig2017light,sato2017exploring}. Others incorporate negative sampling into the estimation method itself \cite{lian2017practical,bian2020learning}. Some datasets include even more user actions besides those of Type 2, such as deletion of stored vacancies from a profile \cite{wang2016analysis,volkovs2017content}, or whether the candidate was hired/rejected \cite{jiang2020learning}, which can be used to define positive/negative instances. 

In case the behavioral data is of Type 1, also all possible pairwise job seeker-vacancy interactions are used. As discussed in Section \ref{sec:jrseval}, this is somewhat less memory demanding as jobs, in this case, are defined at a lower level of granularity \cite{paparrizos2011machine,lin2016machine,heap2014combining}.

\subsection{On the temporal and reciprocal nature of job recommendations}
\label{subsec:temprec}
In \citet{borisyuk2017lijar}, the authors note on interaction with vacancies posted at LinkedIn that \textit{``Since users like to apply for new jobs, our [previous] job recommendation system tends to always show new jobs more often than old jobs”}, a relationship that is also found by \cite{lacic2019should}. Furthermore, what is interesting about the winning RecSys 2016 contribution is that the model explicitly takes into account the recency of items, something which contributions with similar methods (i.e., using XGBoost) did not. 

Whether the claim by Borisyuk et al. (users like to apply to new jobs) holds the question. The relationship could also be reversed: recent vacancies are shown more on top of the recommendation list, and therefore more likely to be clicked. Furthermore, given that vacancies with high demand will find a candidate faster, there is also possibly a survival bias: vacancies with longer lead times are likely to be vacancies with low demand, as those with high demand were already removed from the platform (i.e., as they found a suitable candidate). Nonetheless, irrespectively of the causal chain, we would argue that the vacancy lead time should be taken more into account in job recommender systems, whereas we find that in most literature this aspect is currently neglected.

In one early contribution of job recommender systems \cite{malinowski2006matching}, the authors represent the job recommendation problem as a reciprocal recommendation problem. They state that  \textit{``Theory shows that a good match between persons and jobs needs to consider both, the preferences of the recruiter and the preferences of the candidate”}. Although some contributions do take this reciprocal nature into account, and usually with success (e.g., \cite{borisyuk2017lijar,reusens2018evaluating,le2019towards}), most contributions consider the one-dimensional job seeker perspective, neglecting possible harm to the employer and/or job portal. 

For example, the long tail problem in recommender systems \cite[p. 33]{aggarwal2016recommender} also plays a role in job recommender systems and job search engines (e.g., \cite{de2016predicting}), which in extreme cases may lead to receiving over 1500 applications per vacancy, as was the case for office jobs at the United Nations office in New York \cite[p. 62]{boselie2010strategic}. However, the probability of hiring a candidate seems to be concave increasing in the size of the applicant pool \cite{borisyuk2017lijar}. I.e., growing the size of the applicant pool is likely to lead to diminishing returns. In case there is a large applicant pool for a single vacancy, this not only implies that the recruiter will have to sift through more resumes, but also has to reject more (possibly also well-suited) candidates, which may damage the employer brand. Furthermore, as also shown by \citet{borisyuk2017lijar}, job seekers do apply to less popular but still relevant vacancies, if these are pushed to the top of the recommendation list. 

\subsection{Ethical aspects in job recommender systems}
\label{subsec:ethics}
In their literature review on applied machine learning in human resource management, \citet{strohmeier2013domain} state that \textit{``Ethical and legal aspects are not broadly considered in current research, only equality of treatment and protection of privacy issues are discussed"}. Furthermore: \textit{``A few of the research contributions address equality of treatment [...], mostly by excluding potential discriminatory features from mining"}. Unfortunately, in job recommender literature, equality of treatment is also rarely considered, and if it is considered, the authors indeed come to the conclusion that simply excluding the discriminatory feature would be sufficient (e.g., \cite{qin2018enhancing}).

Excluding discriminatory features may seem logical from the perspective of preserving privacy, or when trying to eliminate human bias, but at the same time this may actually hurt equality of treatment. To check whether the algorithm is discriminating against a certain group, one would require these discriminatory features. Following the argument made by \citet[Ch. 3-6]{perez2019invisible}\footnote{Perez actually provides an argument that this holds more general, Chapters 3 to 6 discuss gender inequality at the workplace.}, the problem of (gender) inequality is perhaps caused by the fact that the impact of HR/recruitment policies, and as we argue here, the choice for implementing a certain job recommender system, are not tested against specific groups. As a result, the chosen policy is often (in Perez' argument male-) biased.

One can easily find examples of how sensible features extracted from a resume or vacancy may lead to inequality of treatment, which is also referred to in the literature as \textit{fairness}. Trivial ones include work experience (age-biased), or first name (gender-biased), but many are also less trivial. These may be the type of words used in a vacancy text \cite[Ch. 4]{berg2018learning}, or examples from O'Neil's well-known \textit{weapons of math destruction}\cite[Ch. 6]{o2016weapons}, of how online (professional) social activity may be gender-biased, or how commuting time may be related to welfare (especially if the office is located in, say, the city center of London). Hence, these should make us aware that it is not an option to assume that if a (discriminatory) feature is not observed, it does not exist. 

The recent increased scientific attention on algorithm fairness \cite{bandy2021problematic} will hopefully turn this tide, as online job portals are more likely to come under scrutiny by algorithm audits. \citet{chen2018investigating} perform such an audit on the candidate search engines of Monster, Indeed, and CareerBuilder, testing these search engines on gender bias. Although all three search engines are found to not allow for direct discrimination on inappropriate demographics (e.g., by allowing to filter on gender), the researchers do find indirect discrimination both in terms of individual and group fairness. \citet{geyik2019fairness}, compare several algorithms for ensuring group fairness in candidate recommender systems/search engines, with a focus on measures for \textit{equal opportunity} and \textit{demographic parity}, one of which has been implemented in LinkedIn. 

Another concern is the usage of so-called ``People aggregators", in which researchers or people from industry use a web crawler to obtain a substantial number of professional or personal profiles. Although this is sometimes with the user's consent (e.g., \cite{diaby2013toward, diaby2014taxonomy, faliagka2014line}), also in many cases this remains unclear. We did not find any work in which user profiles are explicitly anonymized before processing. 

\subsection{JRS at scale: notes from LinkedIn}
\label{subsec:linkedin}
Although many job recommender systems have been proposed in the literature, and the internet holds a considerable number of websites where job seekers can search for jobs, we do find it relevant to put some emphasis on contributions published by LinkedIn employees for a number of reasons. Even though many JRS have been proposed in the literature, few of these dominate the (global) recruitment market \cite{hartwell2015use}. Hence, we consider it likely that job seekers' perception will be biased towards the JRS these larger players are using. 

Furthermore, although many job seekers also use other well-known  general-purpose search engines/social networks, for which the large-scale argument also holds, LinkedIn includes algorithms specifically designed for establishing professional matches. Also, LinkedIn has been considerably transparent about the algorithms they use, and their respective performance, and is capable of testing these algorithms online via their A/B-testing platform XLNT \cite{xu2015infrastructure}.  

LinkedIn's job recommender system/search engine is largely described in \cite{kenthapadi2017personalized, li2016get}, respectively. To facilitate different information needs, LinkedIn encompasses multiple search indices, which are combined into one to facilitate federated search \cite{arya2015personalized}. The job search engine itself is composed of a linear combination of several features, which can be grouped into four classes. First, a query-document semantic matcher. Second, an estimator for the searcher's skill with respect to the set of potential jobs. Third, an estimator for the quality of the document, independent of the query and/or searcher. And fourth, a collection of miscellaneous features such as textual, geographical and social features.

LinkedIn's job recommender system consists of a GLMix model, including a large number of profile features from both the candidate, the job, their interactions, and the cosine similarity between user and job features in the same job domain \cite{zhang2016glmix}. Additionally, the job ranking is adjusted based on the expected number of applications for each job in such a manner that the job seekers are better spread over the set of potential jobs, without reducing the relevance of the high ranked jobs in the recommendation too much \cite{borisyuk2017lijar}.

From considering the different algorithms and their validation, a few observations can be made. Both the JRS and JSE quite heavily rely on the set of skills users indicate on their profile. The algorithms thereby benefit from LinkedIn's policy to nudge users to complete their list of skills, or to provide endorsements to other users, partly by recommending skills to add to their profile \cite{bastian2014linkedin}. Also, of all individual classes of features, the estimate of one's expertise seemed to lead to the largest improvement in model performance. Given that endorsements from other users have a considerable influence on this estimate, the JRS/JSE seems to benefit from LinkedIn being a social network. Although LinkedIn strives towards personalized results, the algorithms are required to be scalable to meet latency requirements \cite{kenthapadi2017personalized}, which leads to some algorithms being competitive to, but not outperform, other models \cite{zhang2016glmix}.

%% file: conclusion.tex
\subsection{Conclusions and directions for further research}
In this paper, we have considered the job recommender system (JRS) literature from several perspectives. These include the influence of data science competitions, the effect of data availability on the choice of method and validation, and ethical considerations in job recommender systems. Furthermore, we branched the large class of hybrid recommender systems to obtain a better view on how these hybrid recommender systems differ. Both this multi-perspective view, and the new taxonomy of hybrid job recommender systems has not been discussed by previous reviews on job recommender systems.

Application-oriented challenges in JRS were already highlighted in early JRS contributions, though, still most literature does not take these into account. Contributions that do take different views on the JRS problem, however, do show that such views can have considerable benefits. These benefits may include improved model performance (temporal perspective), improved distribution of candidates over a set of homogeneous vacancies (reciprocal perspective), or ensuring algorithm fairness (ethical perspective). Currently, most attention goes out to how to represent the substantial amount of textual data from both candidate profiles and vacancies to create job recommendations, for which recently especially deep representations have shown promising results. However, this focus may also create the illusion that this is the only perspective that is relevant. 

Especially in terms of fairness, such a single perspective can be considerably harmful. Although we are not aware of algorithm audits on job recommender systems, an audit on the candidate search engines of Indeed, Careerbuilder, and Monster, did show significant results for both individual and group unfairness in terms of gender. The increased scientific attention towards algorithm fairness, however, does provide algorithms and metrics that can be applied to measure and ensure algorithm fairness. Hence, there is a research opportunity to study how these can be transferred to the job recommender system domain.

Many authors state in the introduction of their contribution that there is a vast amount of data available in the form of vacancies and job seeker profiles. However, there is a clear split in the literature with regards to contributions having also access to interaction data between these two, in particular in the form of clicks/skips on the recommendation list. Interaction data can resolve the language inconsistency between job seekers and recruiters, which is especially troublesome in content-based and some knowledge-based JRS. In case interaction data is missing, one common resort is to use one of the available datasets originating from JRS competitions, in particular the CareerBuilder 2012, RecSys 2016, and RecSys 2017 competitions, which therefore have had a considerable influence on the JRS literature.    

An interesting aspect with respect to the usage of these competition datasets, beyond the contributions to the competitions themselves, is that these datasets are mostly used for training, but rarely for validation. This is unfortunate, as the (to our knowledge) only contribution that compares JRS on different competition datasets shows that error metrics may differ substantially across different datasets. I.e., this raises questions with respect to the generalizability of JRS trained on one dataset. Another interesting question why (online) interaction data is sometimes not taken into account, or along the same line, why researchers often resort to the competition datasets, beyond the motives of contributing the the competition or for validation. Although there may be many valid reasons, we would like to hypothesize from anecdotal experience that it can be difficult to obtain such interaction datasets, as recruitment organizations are not always part of research communities, or given that these recruitment organizations have not always considered the implications of sharing data for research, either from a technical or legal point of view, making it difficult to use such datasets on a short term.

\subsection{Limitations}
Although we believe to have given a broad overview of contributions on job recommender systems, we do like to address some limitations of this review. First, although we managed to further split JRS hybrids into smaller categories, still some classes comprise similar methods. One particular example is that some methods, currently classified as MM-SE, are quite similar to cascade hybrids. I.e., if a cascade hybrid would have used the semantic representation of jobs and job seekers as features in a boosting model, instead of using the similarity between the two as feature, it would have been classified as MM-SE. A similar argument holds for the feature augmentation and cascade hybrid classes.

Second, to limit the scope of this literature review, we only considered literature on job recommender systems. However, such systems do not exist in isolation: they are commonly part of an e-recruitment platform also comprising candidate recommendations, or even employee selection, which are sparsely included in this review.